\documentclass[sigconf]{acmart}
\renewcommand\footnotetextcopyrightpermission[1]{}
\usepackage{amsmath,amsfonts}
\usepackage{algorithmic}
\usepackage{graphicx}
\usepackage{textcomp}
\usepackage{xcolor}
\usepackage{makecell}
\usepackage{pifont}
\usepackage{multirow}
\usepackage{graphicx}
\usepackage{subfigure}
\usepackage{ulem}
\usepackage{color}
\usepackage{booktabs}
\usepackage{url}
\usepackage{tabularx}
\AtBeginDocument{%
	}

\setcopyright{acmcopyright}
\copyrightyear{2018}
\acmYear{2018}
\acmDOI{XXXXXXX.XXXXXXX}

\begin{document}

\title{An Image Dataset for Benchmarking Recommender Systems with Raw Pixels}
\author{Yu Cheng$^{1}$, Yunzhu Pan$^{1}$, Jiaqi Zhang$^{1}$, Yongxin Ni$^{1}$, Aixin Sun$^{2}$, Fajie Yuan$^{1\dagger}$}

\affiliation{
  \institution{$^{1}$Westlake University, $^{2}$Nanyang Technological University}\city{}\country{}
}
\email{{chengyu, yuanfajie}@westlake.edu.cn}

\thanks{$\dagger$ Lead contact. This technical report provides details on the PixelRec dataset and multiple PixelNet baselines. We hope it will stimulate research on image content-based recommendations.}

\renewcommand{\shortauthors}{Cheng et al.}

\begin{abstract}
Personalized recommender systems (RS) have achieved significant success by leveraging explicit identification (ID) features, such as userIDs to represent users, itemIDs to represent items, and various categorical IDs to represent other features. However, the full potential of content features, especially the pure image pixel features, remains relatively unexplored. The limited availability of large, diverse, and content-driven image recommendation datasets has hindered the use of raw images as item representations. In this regard, we present PixelRec, a massive image-centric recommendation dataset that includes approximately 200 million user-image interactions, 30 million users, and 400,000 high-quality cover images. By providing direct access to raw image pixels, PixelRec enables recommendation models to learn item representation directly from them.

    We begin by presenting the results of several classical pure ID-based baseline models, termed IDNet, trained on PixelRec. Then we replace the itemID embeddings (from IDNet) with a robust vision encoder to construct pixel-based baselines, named PixelNet, which exclusively represent items by their raw image pixels. We exhibit the potency and utility of the dataset's image features through a comprehensive study of PixelNet. Our study  establishes benchmark results on  PixelRec, showing the impact of both   recommendation architectures and vision encoders on performance. Furthermore, experiments indicate that even in standard non-cold start recommendation settings, where IDNet is deemed highly effective, PixelNet can achieve equal or even better performance. In addition, PixelNet demonstrates benefits in cross-domain recommendation scenarios via pretraining on PixelRec and increased effectiveness in cold-start scenarios. These outcomes emphasize the significance of visual features in PixelRec. We believe PixelRec can serve as a critical resource and testing ground for research on recommendation models that emphasize image pixel content.
The dataset, code, and leaderboard will be available at \textcolor{blue}{\url{https://github.com/westlake-repl/PixelRec}}.
	\end{abstract}
 
\maketitle
		
	\section{Introduction}

Recommender systems (RS) are a type of artificial intelligence technology designed to deliver personalized item recommendations to users by analyzing their behaviors and preferences. RS has a wide range of applications in various fields, including e-commerce, social networks, online advertising, and more. In the past decade, numerous highly effective recommendation models have been developed. Among them, the explicit ID-based models (referred  to  as IDNet)~\cite{koren2009matrix,rendle2010factorization}, 
which utilize unique IDs to represent users, items, and other category features, exhibit exceptional recommendation efficiency and performance. As a result, IDNet has become the dominant model in the RS community for over a decade~\cite{yuan2023go}.

\textbf{Motivation:} Despite being well-established,  IDNet models still face  some inherent limitations. To be specific, the ID embedding features of IDNet need to be learned from a large amount of user-item interaction data, resulting in poor performance in cold-start scenarios~\cite{he2016vbpr}.  Additionally, IDNet is prone to popularity bias~\cite{abdollahpouri2019managing}, particularly when handling highly popular items. 
 The non-sharable nature of ID features in IDNet also significantly limits its transfer learning capabilities, as shown in ~\cite{ding2021zero,wang2022transrec,hou2022towards}.
 For example, platforms like TikTok and YouTube generally  do not share their userID and videoID data with each other, creating  a significant barrier  for IDNet to achieve a paradigm similar to that in the fields of natural language processing (NLP) and computer vision (CV), where a large pre-trained  \textit{foundation model} can serve various businesses.

  On the other hand, recent advancements in large foundation models, such as BERT~\cite{devlin2018bert}, the GPT series~\cite{radford2018improving,radford2019language,brown2020language}, and various Vision Transformers (ViT)~\cite{DBLP:conf/iclr/DosovitskiyB0WZ21}), have been successful in comprehending multimodal content. Intuitively, these models could provide an alternative way to represent items with rich modality features~\cite{wu2021empowering,xiao2022training,yuan2023go}.When used as item encoders, recommendation models could potentially overcome the limitations of IDNet. For instance, such models have a natural advantage in representing cold items and excel in transfer learning settings due to the generality of modality content features.  This makes cross-domain and cross-platform recommendations more feasible, even with the potential to achieve a universal ``one-for-all" (i.e. foundation) recommendation paradigm~\cite{shin2021one4all,hou2022towards,geng2022recommendation, fu2023exploring}.
Therefore, an active research trend in \textit{modern recommender systems} ~\cite{yuan2023go,Rajput2023gen,wang2022transrec,hou2022learning,singh2023better,ding2021zero,li2023text,li2023exploring} is to \textit{move away}\footnote{There is also a traditional research line 
which  combines  explicit itemID with pre-extracted offline multimodal features  for recommendation~\cite{mcauley2015image, he2016vbpr,he2016ups,du2020learn,sun2019research}, however, as revealed  in~\cite{yuan2023go}, such pre-extracted multimodal features  may not be very useful  for non-cold or warm item recommendation, where itemID embeddings have already been sufficiently trained.
Moreover, once ID features are involved, the key advantages of pure content features may be lost, such as the transfer learning and
 cold-start item recommendation ability~\cite{ding2021zero,hou2022towards,wang2022transrec}.  
} from relying on explicit ID features and instead represent items  using only  their content features, even for non-cold and warm items.

\begin{figure*}[!t]
		\centering
		\includegraphics[width=.75\linewidth]{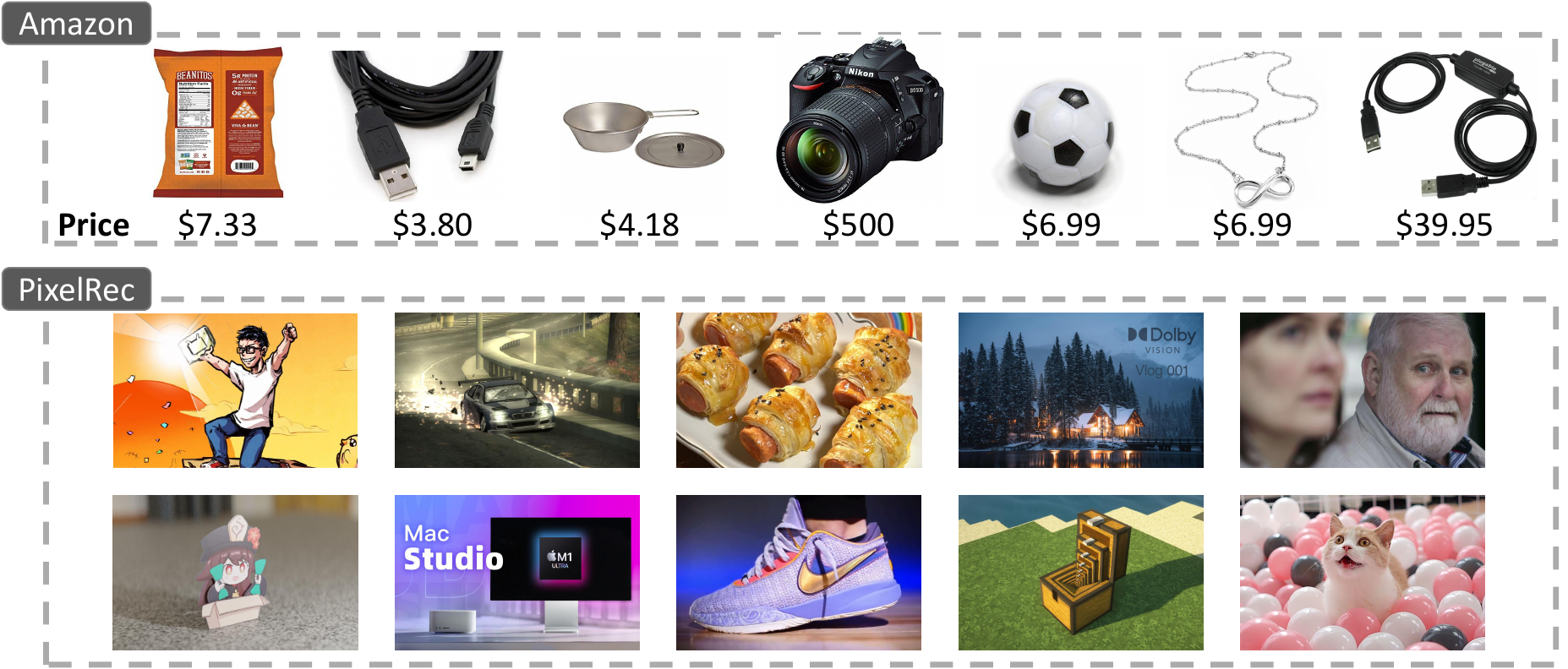}
		\caption{Examples of images in Amazon vs. PixelRec.  Images in PixelRec are more abstract and semantically rich than those in Amazon. Second, most image types found in Amazon (e.g. products), GEST (e.g. food), and HM (e.g. shoes) can also be found in PixelRec. Third, on a leisure and entertainment platform, users' click behavior is more influenced by the content itself, unlike platforms such as Amazon where users' intent-click may be more influenced by factors such as item price and quality for visually similar items.}
		\label{fig:imagecompare}
		
	\end{figure*}

\begin{figure}[t]
	\centering
		\includegraphics[width=\linewidth]{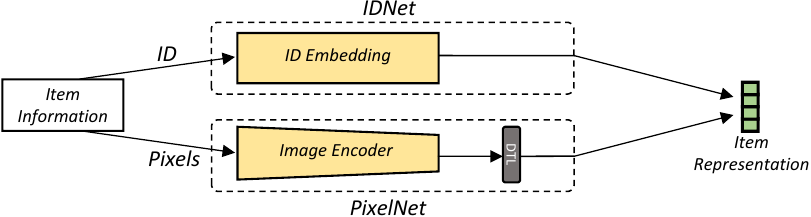}
		\caption{Item representation module of PixelNet and IDNet. DTL is the "Dimension Transformation Layer" with one dense layer. 
  }
		\label{fig:pixelnet}
\end{figure}

\textbf{Challenges:} 
Recently,	a significant amount of related work~\cite{wu2021empowering, hou2022learning,geng2022recommendation,lu2021less} in RS  has conducted research on plain text content recommendation, with many benefiting greatly from the MIND~\cite{wu2020mind} dataset. However,  
few studies have explored learning directly from \textit{raw} pixel features for image recommendation, especially in non-cold item recommendation settings.
One primary reason could be that learning high-level semantic representations of items  from image pixel features is generally more challenging than embedding ID features, especially  several years ago when powerful image encoders (e.g. ResNet~\cite{he2016deep} and ViT) were not yet available.
A second difficulty could be that learning image representations from raw pixels and performing joint optimization of  user and image encoders  is computationally expensive (see Appendix Table~\ref{tab:time_compare}).
In addition to these model-related challenges, the lack of suitable image datasets also poses a challenge for this direction. Specifically,  many publicly available  image recommendation datasets~\cite{he2016vista,wu2019hierarchical}  only include pre-extracted frozen visual features from a specific image encoder without providing access to the raw image pixels. Additionally, user intent in well-known image recommendation datasets, such as  Amazon~\cite{mcauley2015image}, HM\footnote{https://www.kaggle.com/competitions/h-and-m-personalized-fashion-recommendations} and GEST (Google restaurants)~\cite{yan2022personalized} is heavily influenced by non-visual features (refer to Figure~\ref{fig:imagecompare})  such as item price, sales, brand, and delivery location. As a result, learning item representations  from raw pixel features alone (refer to Figure~\ref{fig:pixelnet}) may not lead to satisfactory results in these e-commerce datasets.

\textbf{Proposal:} Motivated by the above challenges, we aim to promote  research on image recommendation by directly representing images  using their \textit{raw} pixels, rather than relying on or being assisted by itemID features (refer to \textit{Footnote} 1). To facilitate such research, we  release a visual content-driven image recommendation dataset.
Specifically, we introduce PixelRec
, a very large, diverse, and high-quality dataset of cover images collected from a content-only recommender system. PixelRec contains approximately 200 million user image interactions, 30 million users, and 400,000 high-resolution micro-video cover images. By providing access to raw image features, PixelRec enables recommendation models to 
learn item representation directly from them.
To demonstrate its utility, we conduct empirical and benchmark studies. First, we evaluate several representative IDNet baselines. Then, 
we introduce PixelNet, which substitutes  the ID embedding component  of IDNet with a \textit{trainable} modern vision encoder that represents items using raw image pixel features (see Figure~\ref{fig:pixelnet}).
  Our findings demonstrate that PixelNet, even in standard non-cold start recommendation settings, can perform on par or better than their IDNet counterparts  with a sequential recommendation backbone.
Furthermore, we study PixelRec in the conventional  cold-start and cross-platform recommendation settings,   where PixelNet has a clear advantage over IDNet. 
	Our comprehensive results demonstrate the importance of utilizing raw visual features for recommendation and highlight the significance of PixelRec in this area.

	\section{PixelRec Dataset}
 		\begin{figure}[t]
		\centering
		\includegraphics[width=1.0\linewidth]{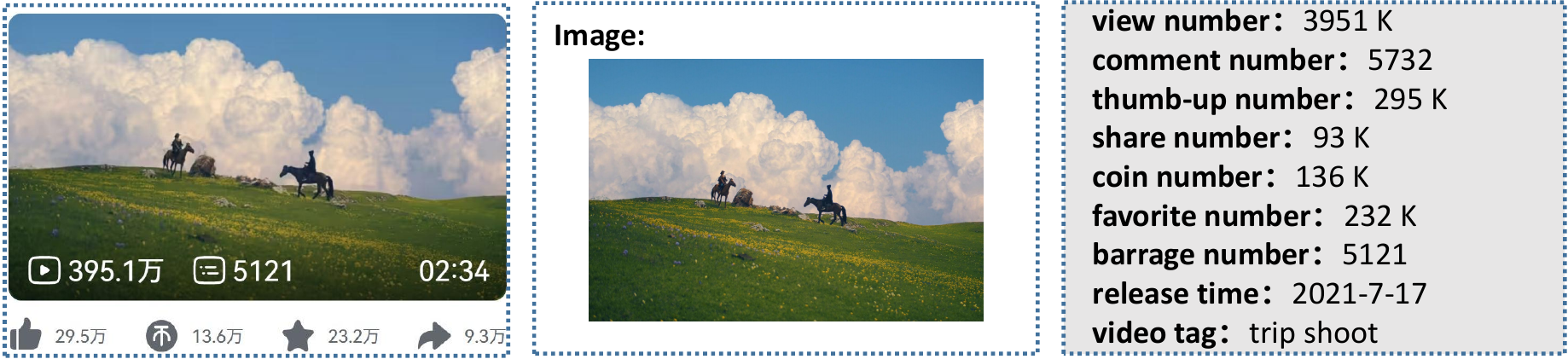}
		\caption{An example of a video cover image and other features in PixelRec.}
		\label{fig:sample}
		
	\end{figure}

 In this section, we first describe the data collection process, then compare PixelRec with existing image recommendation datasets, and highlight its unique characteristics.
	\subsection{Dataset Curation}
	\label{datasetcollect}
	All data in PixelRec was collected from an online video sharing website 
 with a focus on content consumption  rather than e-commerce, unlike websites such as Amazon, H\&M, and GEST. The entire data crawling process took 13 months, from September 2021 to October 2022. In order to ensure item diversity, we  collected items\footnote{For this study, we focus only on micro-videos, which are defined as videos with a playback time of less than 10 minutes.} from  both main channel (with various video categories) and 22 different vertical channels (including technology, cartoons, games, movies, food, fashion, sports, etc.). 
 We use  the cover image to represent a video instead of using the original video itself. This is mainly because the cover image of a micro-video serves as the user's first impression of the video and  has been found  to have a strong correlation with their video click behavior, as reported in \cite{koh2022exploration,lee2017large,deldjoo2016content}. Notably, on the video watching platform, clicks on videos are almost equivalent to clicks on cover images, as the cover image occupies more than 95\% of the click trigger range. 

By requesting page updates, we were able to obtain new video links from the home page of each channel. After dozens of requests, we collected between  1000-5000 videos per channel. Then, we  visited  the  pages of these collected videos, which usually contained a large number of external links to other videos. From each page, we randomly selected 1-3 videos and repeated this process several times. Then, we merged all items and removed any duplicates.
In addition to the cover images, we also collected many other video features (see Figure~\ref{fig:sample}) for supporting other research, including \{view\_number, comment\_number, thumb-up\_number, share\_number, coin\_number, favorite\_number,  barrage\_number, release time\}, text contents: \{title, description\}, and a video tag.

After that, we proceeded  to collect user feedback, which is the core element in a recommendation dataset. We visited all pages of  collected videos and retrieved user comments (including bullet comments) and timestamps. We opted to collect comments instead of clicks for two reasons. Firstly, comments can be considered  stronger\footnote{One may argue that a negative comment means the user didn't like the item, which is true for the e-commerce dataset, but not for our PixelRec, since most comments are about the user feeling towards  the video story rather than whether it is worth clicking or not.} click signals because users need to  click on the cover image to enter the video watching page before they can leave a comment there (i.e. \textbf{comments $\Rightarrow$  clicks on cover images}). Secondly, comment data is publicly available and can be crawled without privacy concerns. In contrast, obtaining user click data requires access to the company's servers, which raises concerns regarding privacy and data security. Note that we only gathered userID information from the comments, rather than the actual comment content.
 For each video, we recorded up to 6500 interactions and ignored multiple interactions from the same user. 
Finally, we merged and aggregated all users, items, and corresponding user-item interaction behaviors.

In total, PixelRec contains 200 million user  actions with about 30 million users and 400,000 high-quality  cover images. We make the entire dataset of PixelRec publicly available. However, it is difficult for academic teams  to experiment on such a huge dataset, especially for PixelNet algorithms  trained jointly with a large image encoder. 
 To this end, we provide several PixelRec versions through sampling, namely \textbf{Pixel200K}, \textbf{Pixel1M}, and \textbf{Pixel8M}. 
 We first randomly sampled 1 million users with at least 5 comments to  build Pixel1M. Then, we randomly selected 200,000 users and their actions from Pixel1M to build Pixel200K,  a medium-sized dataset mainly for research purposes. At last, we collected all users who have at least  5 comments to construct Pixel8M  (with about 8 million users).
 Next, we first show how PixelRec differs from existing visual recommendation datasets and then provide basic statistics of it.
	\subsection{PixelRec vs. Related Datasets}
 \label{relatedwork}

 	\begin{table}[t]
		\centering
		\caption{Image RS datasets. MM is the pure multimedia scene, where user preference is mainly determined by item content.}
	\begin{tabular}{lc c c c  }
			\toprule
			& \multirow{3}{*}{\makecell[c]{ Flicker \\ Behance}} &  \multirow{3}{*}{\makecell[c]{Pinterest \\ WikiMedia}}   &   \multirow{2}{*}{\makecell[c]{GEST \\ HM}} &  \multirow{3}{*}{PixelRec} \\
			& &  & &  \\
			& &  & Amazon& \\
			\midrule
			Raw Image & \ding{56} & \ding{52} & \ding{52} & \ding{52}  \\
			Large Scale & \ding{56} & \ding{56}  &\ding{52}  &  \ding{52} \\
			Scene & MM & MM & e-commerce &    MM \\
			\bottomrule
		\end{tabular}
		
		\label{tab:relatework}
	\end{table}	

	There are several commonly used image recommendation datasets, summarized in Table ~\ref{tab:relatework}.\footnote{Note that  we do not discuss  multimodal datasets if they have no user interaction data, such as YouTube-8M~\cite{abu2016youtube}. } Most such datasets, such as  Behance~\cite{he2016vista} and Flicker~\cite{wu2019hierarchical},  only provide frozen  visual  features  extracted from pre-trained encoder networks, e.g. AlexNet~\cite{krizhevsky2017imagenet} and ResNet. However, these invariant features learned  from  the computer vision tasks are not optimal for  recommendation tasks~\cite{yuan2023go}, and thus, recommendation models that rely solely on them may not perform satisfactorily (see Section~\ref{E2EEvaluation}).  Instead, with PixelRec, we are able to build recommendation models that learn directly from  raw image pixels using  trainable image encoders. Pinterest~\cite{geng2015learning} and ~WikiMedia~\cite{parra2021visrec} are two popular  datasets for visual recommendation with raw image pixels, but both of them are   small in size, which cannot meet the requirements of the increasing scale of recommendation research.
	
 
Amazon, HM, and GEST are three large-scale e-commerce datasets with raw images. Prior works on visual recommendation \cite{he2016vbpr,he2016ups,du2020learn} have extensively explored these datasets, where visual features are often used in combination with other non-visual features. This is because  user's purchase or click behavior on Amazon, HM,  and GEST is heavily  influenced by various critical factors, such as item price,  sales, brand, seller reputation, proximity, and more importantly, user's current demand  or query, which cannot be achieved merely  by learning appearance features.  
In addition, the themes of their pictures are relatively restricted,  with the majority of the features being either individual  products (e.g. in Amazon \&  HM) or  belonging to the same category (e.g. mainly about food and restaurants in GEST).
 Therefore, they may not be representative  for semantically  more diverse and complex image recommendation tasks (see Figure~\ref{fig:imagecompare}).
In contrast, on a content-focused platform rather than an e-commerce platform,  image characteristics or visual features play a more prominent role in attracting user interaction.
Our assumption in PixelRec is that \textbf{if a user leaves a comment, it can be reasonably assumed that they have clicked on the video's cover image beforehand (i.e. comments $\Rightarrow$  interactions on cover images). } This is almost the only way to leave comments on the  platform, as manually entering the video watching URL is not a common practice for most users. Note we cannot assume the opposite --- items without comments are not preferred ---  to be true, which however is a common property in most recommendation datasets, i.e. items without interaction can be either positive or negative for the user, also known as the one-class recommendation problem.

\subsection{Analysis of PixelRec} 
\begin{figure}[t]
	\centering		\includegraphics[width=1\linewidth]{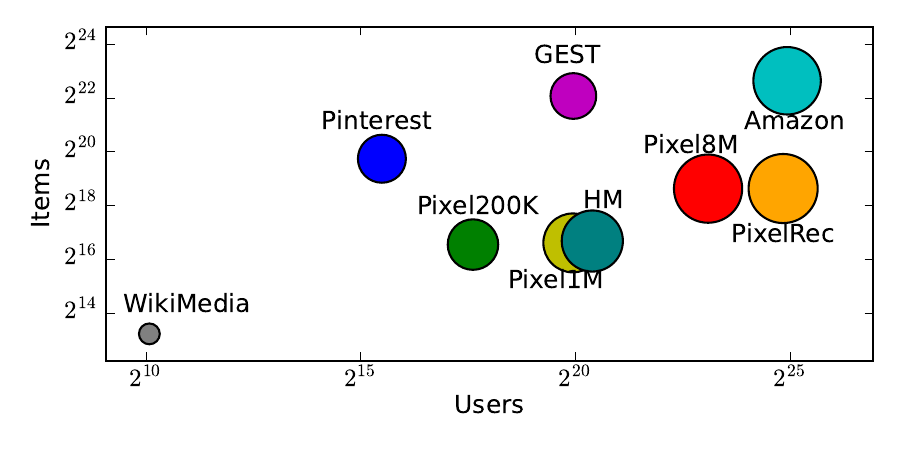}
 \vspace{-0.8cm}
		\caption{Dataset size.
			Each circle represents a dataset. The x-axis is the number of users, and the y-axis is  the number of items. The area of a circle represents the number of user-item interactions in logarithm scale.}
		\label{fig:scale}
\end{figure}

We believe PixelRec is a valuable addition to the existing image RS datasets, exhibiting the following key properties.
\begin{itemize}
	\item Having raw content:  as shown  in Figure~\ref{fig:imagecompare} and~\ref{fig:sample},  PixelRec  offers researchers access to a vast image dataset that consists of high-resolution raw image pixels. This feature enables recommendation models to represent items directly using the raw pixel features, which can lead to more effective and precise image-based recommendations.
	\item Having rich features: 
	as shown in Section~\ref{datasetcollect}, PixelRec offers  a plethora of  features for evaluating the click-through rate (CTR) prediction task, although this is not the primary focus of this paper.  
 \item Diversity: 
PixelRec includes 118 tags  that encompass a broad range of topics, such as movie clips, games, variety shows, food, entertainment stars, daily life, cartoons, technology, and more.
	It  encompasses most image types in HM, GEST, and Amazon. Furthermore, the item images in PixelRec are more complex and contain greater semantic content compared to those in Amazon and HM, where the items are primarily single products that require less knowledge to understand, see Figure~\ref{fig:imagecompare}.  For this reason, PixelRec can serve as a more challenging   benchmark dataset for image recommendation.
	\item Large scale: as shown in Figure~\ref{fig:scale}, PixelRec is among the largest visual recommendation datasets, featuring the highest volume of user-image interaction data.\footnote{
		For Amazon dataset statistics, we have removed items without image urls. Also note that while Pinterest has about 880K (K is thousand) items, 780K of them have interactions less than 6. Besides, it has only less than 40K users.}
Given its diversity and large scale, PixelRec holds promise to be used as a valuable pre-training dataset for establishing \textit{foundation} image recommendation 
models~\cite{bommasani2021opportunities}.

\item Content-driven recommendation scenario: The user interaction data in PixelRec comes from a leisure and entertainment-oriented content platform, which is significantly different from that of e-commerce platforms. Therefore, it has the potential to be a valuable dataset for research on building recommendation models that emphasize item content.

\end{itemize}

\begin{figure}[t]
		\centering
		\subfigure[Pixel1M]{
			\begin{minipage}[h]{0.225\textwidth}
				\includegraphics[width=1\textwidth]{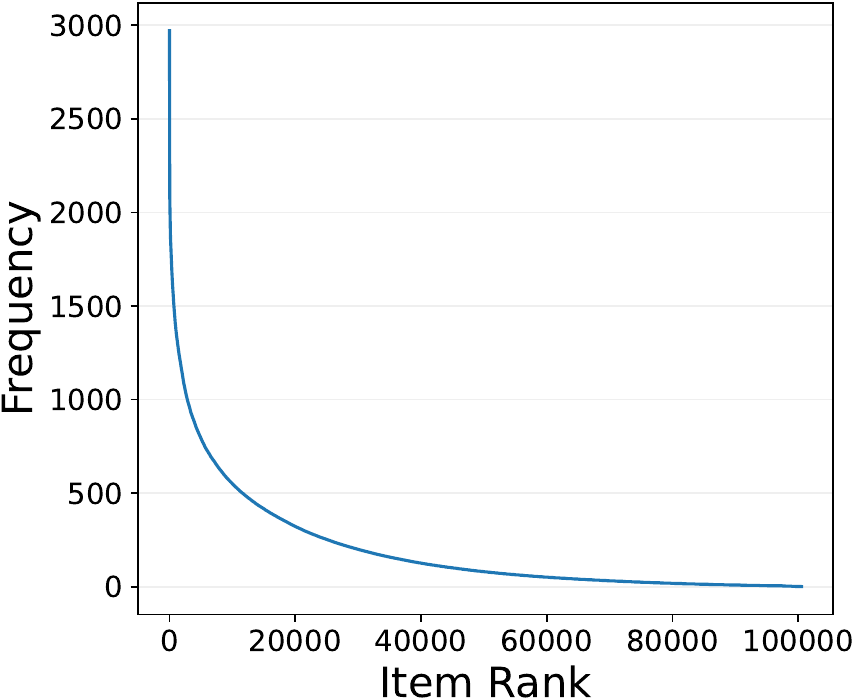}
                \hfill
			\end{minipage}
		}
		\subfigure[PixelRec]{
			\begin{minipage}[h]{0.225\textwidth}
				\includegraphics[width=1\textwidth]{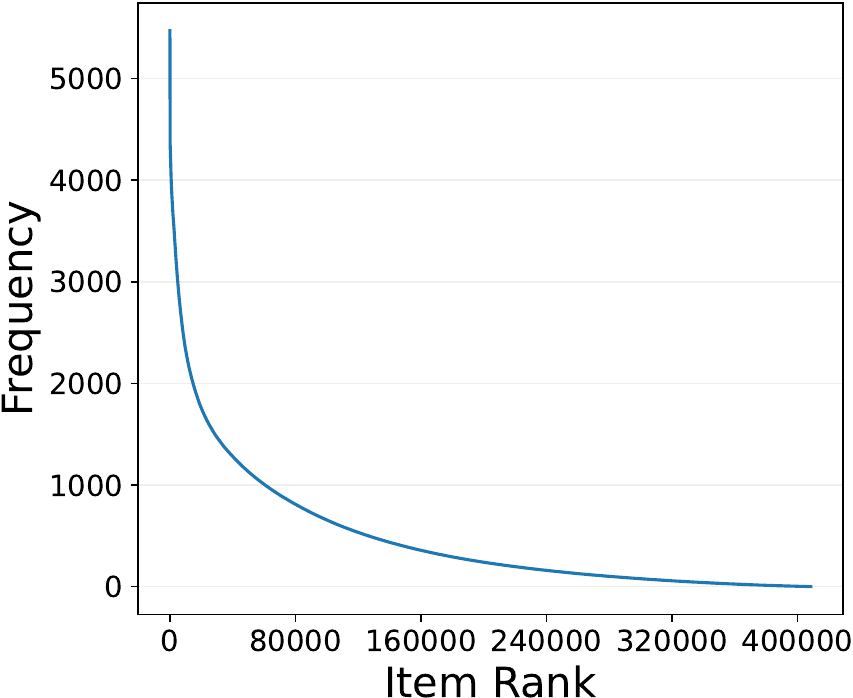}
                \hfill
			\end{minipage}
		}
		\caption{Item popularity distributions of Pixel1M (left) \& original PixelRec (right).}
  
		\label{fig:Ldistfigure}
	\end{figure}

\begin{figure}[t]
		\centering
		\subfigure[Item Rank]{
			\begin{minipage}[t]{0.22\textwidth}
				\vspace{0pt}
				\includegraphics[width=1\textwidth]{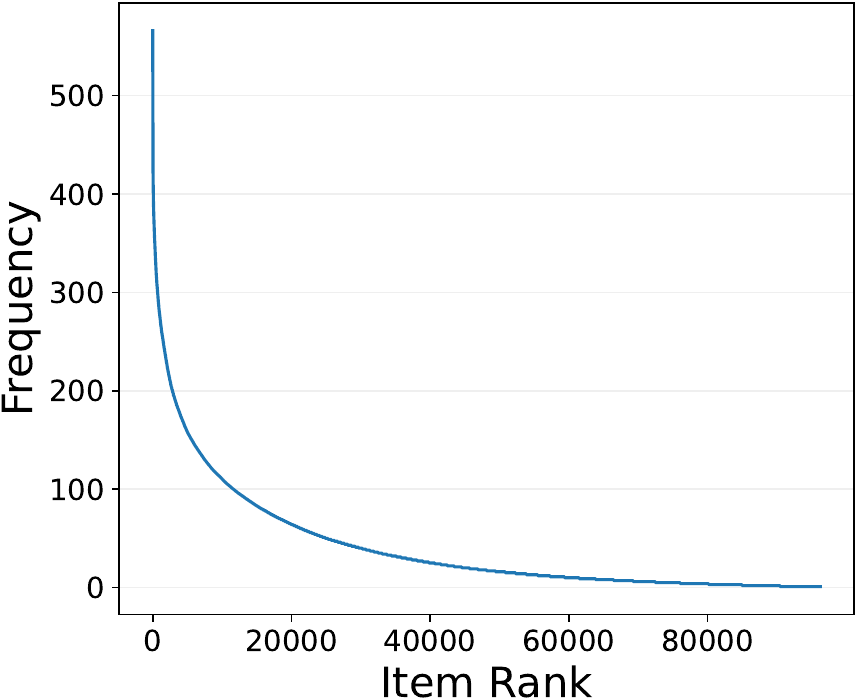}
				
			\end{minipage}
			\label{fig:itemrank}
		}
		\subfigure[Session Length]{
			\begin{minipage}[t]{0.22\textwidth}
				\vspace{0pt}
				\includegraphics[width=1\textwidth]{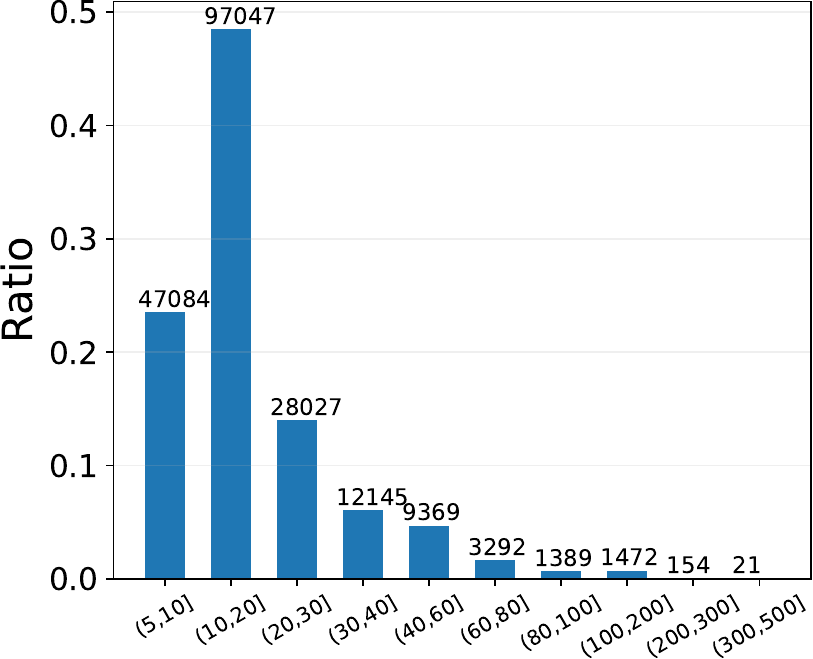}
				
			\end{minipage}
			\label{fig:sessionlen}
		}
		\caption{Data distribution of Pixel200K. (a) is item popularity distribution,  (b) is session length histogram of user behavior. }
		\label{fig:distfigure}
	\end{figure}

As mentioned above, we provide three sub-datasets by sampling, i.e. Pixel200K, Pixel1M, and Pixel8M.  we  report statistics of them in Table~\ref{tab:ltik} and Figure~\ref{fig:Ldistfigure}, and more details of
Pixel200K in Table~\ref{tab:tikdata} and Figure~\ref{fig:distfigure}.

\textbf{Pixel200K statistics:} As shown, Pixel200K contains 200,000 users, 96,282 cover images and 3,965,656 user image interactions. The average interaction per user and per item is 19.83  and 41.19 respectively.  
Figure~\ref{fig:itemrank} shows that item popularity where Pixel200K follows a typical long-tail distribution, with the most popular item counting 566 times and the coldest item appearing only once.  Figure~\ref{fig:sessionlen} shows that half of  user behavior sessions are between 10 and 20 in length. The longest behavior session length is  434 and the shortest is~6. Other statistics of Pixel1M, Pixel8M, and PixelRec are shown in our github page.

	\begin{table}[t]
		\centering
		\caption{Statistics of Pixel1M, Pixel8M and original PixelRec.}
		\label{tab:ltik}
		\begin{tabular}{lrrr}
			\toprule
			& Pixel1M    &Pixel8M  & PixelRec    \\
			\midrule
			\#User &  1,001,822   &   8,886,078      &29,845,039\\
			
			\#Item  & 100,541    &  407,082        &408,374  \\
			
			\#Interaction &  19,886,579  & 158,488,652 &195,755,320  \\		
			\bottomrule
		\end{tabular}
	\end{table}
	\begin{table}[t]
		\centering
		\caption{Statistics of Pixel200K. \#Inter. denotes the number of interactions. \#User.avg denotes the average length of user behavior sequence; \#Item.avg denotes the average number of user interactions per item. }
		\label{tab:tikdata}
		\centering
		\begin{tabular}{l p{0.9cm}<{\centering} l p{0.9cm}<{\centering} l p{1cm}<{\centering}}
			\toprule
			\#User &  200,000   &\#Item  & 96,282 &\#Inter. &  3,965,656 \\
			\#User.avg  & 19.83 &\#Item.avg & 41.19 & Sparsity & 99.97\%  \\
			\bottomrule
		\end{tabular}
	\end{table}

\subsection{Privacy and Copyrights}
PixelRec only includes public user behaviors for privacy protection. Both userIDs and itemIDs have been anonymized.
To avoid copyright issues, we provide image URLs and  a special tool to permanently  access and download related image data. This is a common practice in most previous literature~\cite{zeng2022tencent,nielsen2022mumin}
when publishing multimedia datasets.
We will also provide the original dataset for download under the ImageNet license, following https://www.image-net.org/download.php.

\section{Experimental Setup}
\subsection{Evaluation}
We adopt the popular leave-one-out strategy to split PixelRec into training, validation, and testing sets. That is, the last interaction of each user is used for testing, the penultimate one for validation, and the rest for training. After  splitting the data on Pixel200K, the training set contains 95,031 items and 3,565,656 interactions, the validation set contains 46,546 items, and the test set contains 40,398 items. 
We apply two popular top-N metrics, i.e. Recall@N and NDCG@N (Normalized Discounted Cumulative Gain), to  evaluate recommendation performance. 
N is set to 5 and 10. Note that  we rank
the predicted item among all items in the pool instead of
drawing 100 random items. 
\subsection{Baselines and Training}
  In this paper, we primarily study the performance of two  classes of baselines on PixelRec.
The first category is the traditional baselines, including pure
ID neural models (IDNet) and visual-aware recommendation models (ViNet) that apply pre-extracted frozen visual features with/without ID features.
The second category is PixelNet, which replaces ID embeddings with modern vision encoders that learn  directly from raw image pixels (see Figure~\ref{fig:pixelnet}).
PixelNet jointly optimizes the recommendation backbone and vision encoders, and produces notably better results than using  pre-extracted invariant visual features (refer to Section ~\ref{E2EEvaluation}). But it also requires considerable computational resources and training time (see Appendix Table~\ref{tab:time_compare}).\footnote{Note this does not affect the latency of online service, as visual features can be pre-computed in the offline stage.} Therefore, PixelNet has received little attention in the literature so far.
Nonetheless, we are optimistic that, with PixelRec, researchers in our community could develop more practical and advanced PixelNet models in the future,  even beyond the mainstream PixelNet and ViNet paradigms.\footnote{Just recently, we have noticed that there may be another potential direction: representing raw images using semantic IDs~\cite{Rajput2023gen,petrov2023generative}. This approach has the potential to effectively mitigate the computation issues, but its effectiveness requires further study.
Another very recent study~\cite{yang2022gram} also proposed a highly promising solution to improve the training efficiency, which appears to be feasible for PixelNet.
} 
Without special mentioning, all models in this paper are based on the popular pairwise BPR~\cite{rendle2012bpr} loss and random negative item sampling.

\subsubsection{IDNet}
We report benchmark  results for two  types of IDNet baselines. The first type is the typical factorization models that do not explicitly model user sequential patterns.

\begin{itemize}
	\item \texttt{MF}~\cite{rendle2012bpr}. It factorizes the sparse user-item interaction matrix into two dense smaller matrices to represent user embeddings and item embeddings.
	\item \texttt{DSSM}~\cite{huang2013learning}. It uses two standard DNN encoders to model user and item features respectively. Unlike MF, we represent a user by their interacted itemIDs instead of their userIDs.
	
	\item \texttt{FM}~\cite{rendle2010factorization}. It  models  the explicit interaction between every two features. Like DSSM, we  use the  past interacted items to represent the user and the itemID to represent target item.	
 
	\item \texttt{LightGCN}~\cite{he2020lightgcn}. It is a simplified graph convolutional network (GCN) by linearly propagating user and item embeddings on their interaction graph.

\end{itemize}

The second type is based on sequential network architecture that learns user preference from the user-item interaction sequence. We report results on several most popular network backbones, namely RNN-based \texttt{GRU4Rec}~\cite{hidasi2015session},  CNN-based 
\texttt{NextItNet}~\cite{yuan2019simple}, GNN-based \texttt{SRGNN}~\cite{wu2019session}, 
multi-head self-attention (MHSA) based   \texttt{SASRec} \cite{kang2018self}, \texttt{BERT4Rec}~\cite{sun2019bert4rec} and  \texttt{LightSANs}\cite{fan2021lighter}. For the sequential recommendation task, we split the entire user behavior sequence in the training set into multiple subsequences of length 10. The validation and testing sets are  the same as before.

\subsubsection{ViNet}
We evaluate three visual methods, all of them use frozen visual features pre-extracted from pre-trained image encoders (i.e. ResNet50 pre-trained in CLIP~\cite{radford2021learning}):
\begin{itemize}
	\item \texttt{VisRank}.  A simple baseline model introduced in~\cite{kang2017visually}, which makes recommendations by exploiting
	similarity of content features.
	
	\item \texttt{ACF}~\cite{chen2017attentive}. It adopts the attention mechanism to learn user preference  with both item-level and component-level (e.g. regions in an image) \textit{implicitness}. 
	
	\item \texttt{VBPR}~\cite{he2016vbpr}. It is an extension of MF which adds visual features as a complement to the itemID embeddings.  
\end{itemize}

\subsubsection{PixelNet}
PixelNet apply the exact same network backbone, loss function, and negative sampler as IDNet by replacing the itemID embedding component with image vision encoders. 
In contrast to ViNet, PixelNet jointly  trains both user encoder and image encoder(s).
Note that LightGCN cannot be used directly for this purpose since message passing over a large number of items is unfeasible with  joint learning.

\subsection{Detailed Settings}

\subsubsection{Data Processing}
Let $\mathbf{u}$, $\mathbf{i}$, $\mathbf{t}$ denote the user, item and timestamp of interactions. 
Let $\mathbf{I_u}$ denote the behavior sequence of user $\mathbf{u}$, which is an ordered  list of items sorted by interaction time: [$\mathbf{i_1}$, $\mathbf{i_2}$ ... $\mathbf{i_{N_u}}$], where $N_u$ is the behavior sequence length of  $\mathbf{u}$. Let $n$ denote the maximum sequence length of the recommendation model, and the behavior sequence will be split into one or multiple fixed-length subsequences, e.g. the first one is [$\mathbf{i_1}$, $\mathbf{i_2}$ ... $\mathbf{i_n}$], $n$ is set to 10 in our experiment. 
The data format is processed according to the requirement of the respective algorithms, as shown in Table ~\ref{tab:dataprocess}.

\begin{table}[t]
	\centering
	\caption{Formulation of different baselines. For example, $\mathbf{u} \rightarrow \mathbf{i}$  denotes that these models aim to predict target $\mathbf{i}$ for user $u$. Item $\mathbf{i}$ can be represented by itemID in IDNet, pre-extracted features (in ViNet), or an image encoder in PixelNet.}	
	\label{tab:dataprocess}
		\begin{tabular}{ l l }
			\toprule
			Model &  Formulation   \\
			\midrule
			MF, VBPR,   LightGCN &  $\mathbf{u} \rightarrow\mathbf{i} $; \\
			
			\midrule
			\multirow{3}{*}{DSSM, FM} & $\mathbf{i_1}, \mathbf{i_2} ...\mathbf{i_{n-1}} \rightarrow \mathbf{i_n} $;    \\
			& ...... \\
			& $\mathbf{i_2}, \mathbf{i_3} ...\mathbf{i_{n}} \rightarrow \mathbf{i_1} $; \\
			\midrule
			\multirow{3}{*}{ACF}& $\mathbf{u}, \mathbf{i_1}, \mathbf{i_2} ...\mathbf{i_{n-1}} \rightarrow \mathbf{i_n} $; \\
			& ...... \\
			& $\mathbf{u}, \mathbf{i_2}, \mathbf{i_3} ...\mathbf{i_{n}} \rightarrow \mathbf{i_1} $;\\
			\midrule
			GRU4Rec, NextItNet, SASRec   & $\mathbf{i_1}, \mathbf{i_2} ... \mathbf{i_{n-1}} \rightarrow \mathbf{i_2}, \mathbf{i_3}...\mathbf{i_n} $; \\
			\midrule
			BERT4Rec&  $\mathbf{i_1}, [\text{MASK}], ...\mathbf{i_{n}} \rightarrow  \mathbf{i_2}$; \\
			
			\midrule
			\multirow{3}{*}{SRGNN, LightSANs} &  $\mathbf{i_1}, \mathbf{i_2}$ $\rightarrow \mathbf{i_3} $; \ \  \  $\mathbf{i_1}, \mathbf{i_2}, \mathbf{i_3}$ $\rightarrow \mathbf{i_4} $;  \\
			& ...... \\
			& $\mathbf{i_1}, \mathbf{i_2} ...\mathbf{i_{n-1}} \rightarrow \mathbf{i_n} $; \\
			\midrule
			VisRank&  $\mathbf{i_1}, \mathbf{i_2} ...\mathbf{i_{N_u-1}} \rightarrow \mathbf{i_{N_u}} $; \\
			
			\bottomrule
		\end{tabular}
\end{table}	

\subsubsection{Hyper-parameter Setting}
In order to establish a strong baseline on PixelRec, we exhaustively search the hyper-parameters of traditional baseline models, i.e. IDNet and ViNet,  on the validation set. For example, we search the embedding size  for IDNet in  [128, 512, 1024, 2048, 4096, 8192], where we surprisingly find that 
the best value can be 4096 for some IDNet (i.e., MF, DSSM, and FM).
For learning rate,  we search them in [1e-6,5e-5,...,1e-3]. 
However, conducting such an extensive hyper-parameter search for PixelNet is infeasible, as it typically requires significantly longer training time than IDNet. Therefore, we searched them around the best values found for IDNet. We believe that finding an efficient method to discover the optimal hyper-parameters for PixelNet is an important but unexplored challenge for the research community.

\subsubsection{Transfer Learning Setting}
To conduct the effects of transfer learning, we choose PixelRec  as the source domain dataset. As for the target domain dataset, we crawled a small dataset from Tencent-News\footnote{https://news.qq.com/} following a similar strategy to PixelRec, where the items are micro-videos and we still represent them with their cover images. The target dataset contains 20K  users, 3.7K items and 144K interactions. We will first pre-train PixelNet on PixelRec, then fine-tune it on the target dataset. To facilitate  future research, we will also release this smaller target dataset.

\section{Traditional Benchmark}
\label{TraditionalBenchmark}

We report main experimental results of traditional baselines (IDNet and ViNet) on Pixel200K in Table~\ref{tab:benchmark}, and partial results on the larger Pixel1M in Appendix Table~\ref{tab:benchmark_large}.
First,  it is evident that sequential IDNet models are generally more powerful than these non-sequential models. For instance, SASRec, LightSANs, and NextItNet exhibit significantly higher accuracy than MF and LightGCN. 
It might have two reasons: 1) models that represent users as  a series of item interactions are more effective than   representing them by their unique userIDs (refer to Table~\ref{tab:dataprocess}), e.g. MF vs. DSSM, MF vs. FM, and MF vs. GRU4Rec;
2) the sequential  backbone (e.g. RNN, CNN, MHSA) of IDNet is more expressive than typical DNN or FM model when modeling  user sequence data. Similar observations have been frequently reported  in  prior literature~\cite{kang2018self,sun2019bert4rec}.  Note that due to GPU memory issues, we can only set the  embedding size of LightGCN to 256, which is much smaller than 4096 in MF, FM, and DSSM.

Second, the ViNet models (VBPR, ACF, and VisRank) perform worse than basic IDNet. For example, VBPR performs worse than MF even though it incorporates  additional visual features.  In fact, we note in the literature~\cite{du2020learn,tang2019adversarial} that  ViNet exhibits better performance than IDNet mainly in the cold start setting. 
Pre-extracted visual features might not be very useful (or even become noise) in regular non-cold recommendation settings, especially for well-trained popular items, also see~\cite{yuan2023go}.

\begin{table}[t]
	\caption{Traditional  baseline (IDNet \& ViNet) results (\%) on Pixel200K. We include a random  baseline and a popularity-based baseline (Popularity) for reference. It should be noted that the reason for the lower absolute value is simply  due to the very large candidate item size and
smaller recommendation list size (e.g. 5 or 10).} 
	\label{tab:benchmark}
        \begin{center}
	\begin{tabular}{l c c c c }
		\toprule
		Method  & Recall@5 &  NDCG@5 & Recall@10 &  NDCG@10    \\
		\midrule
		Random &0.006 &0.003 &0.010 &0.004 \\
		Pop &0.035 & 0.019&0.066 & 0.029\\
		\midrule
		LightGCN & 0.538 & 0.337 & 0.929 & 0.461\\
		MF & 0.558 & 0.344 & 1.013 & 0.490\\
		DSSM & 0.840 & 0.522 & 1.401 & 0.701\\
		FM & 0.816 & 0.506 & 1.357 & 0.679\\
		\midrule
		SRGNN & 0.953 & 0.602 & 1.597 & 0.808\\
		GRU4Rec & 1.094 & 0.700 & 1.833 & 0.937 \\
		BERT4Rec & 1.154 & 0.732 & 1.972 & 0.994 \\
		NextItNet & 1.396 & 0.899 & 2.187 & 1.153 \\
		SASRec & 1.640 & 1.074 & 2.500 & 1.350 \\			
		LightSANs &1.651 & 1.087&2.578 &1.384 \\
		\midrule
		VisRank & 0.226 & 0.154 & 0.347 & 0.193 \\
		ACF & 0.441 & 0.276 & 0.758 & 0.377\\
		VBPR & 0.467 & 0.288 & 0.832 & 0.405\\	
		\bottomrule
	\end{tabular}
        \end{center}
\end{table}

\section{PixelNet Results}

\begin{table*}[t]
	 	\centering
	 	\caption{Performance(\%) of PixelNet on Pixel200K. Figure~\ref{imeben} displays results of several significantly larger vision encoders.}
	 	\label{tab:compare}
	 	\begin{tabular}{p{1.3cm}<{\centering} p{1.3cm}<{\centering} p{1.1cm}<{\centering} p{1.1cm}<{\centering} p{1.1cm}<{\centering} p{1.2cm}<{\centering} p{1.2cm}<{\centering} p{1.2cm}<{\centering} p{1.2cm}<{\centering} p{1.2cm}<{\centering} p{1.2cm}<{\centering} }
		 		\toprule
		 \multirow{2}{*}{ImgEnc} &\multirow{2}{*}{Metrics} & \multicolumn{3}{c}{Non-Sequential Recommender} & \multicolumn{6}{c}{Sequential Recommender}  \\
		 		\cmidrule(r){3-5} \cmidrule(r){6-11}
		 		& &   MF &  FM & DSSM & SRGNN & GRU4Rec & BERT4Rec & NextItNet & SASRec & LightSANs \\
		 		\midrule
		 		\multirow{4}{*}{RN50} 
		 		& Recall@5  & 0.203   & 0.593   & 0.567   & 1.349 & 1.343   & 1.404   & 1.270   & 1.546   & 1.461 \\
		 		& NDCG@5 & 0.120   & 0.363   & 0.349   & 0.852 & 0.832   & 0.883   & 0.794   & 0.972   & 0.919 \\
		 		& Recall@10  & 0.357   & 1.024   & 0.960   & 2.224 & 2.294   & 2.391   & 2.140   & 2.633   & 2.417 \\
		 		& NDCG@10  & 0.169   & 0.501   & 0.475   & 1.132 & 1.138   & 1.199   & 1.073   & 1.321   & 1.226 \\
		 		\midrule
		 		\multirow{4}{*}{ViT}
		 		&  Recall@5  & 0.264   & 0.632   & 0.712   & 1.261 & 1.210   & 1.431   & 1.280   & 1.522   & 1.443 \\
		 		&  NDCG@5  & 0.162   & 0.385   & 0.448   & 0.779 & 0.745   & 0.903   & 0.807   & 0.952   & 0.898 \\
		 		&  Recall@10  & 0.472   & 1.124   & 1.242   & 2.152 & 2.102   & 2.450   & 2.215   & 2.583   & 2.461 \\
		 		& NDCG@10 & 0.229   & 0.543   & 0.617   & 1.065 & 1.031   & 1.230   & 1.106   & 1.292 & 1.224 \\
\bottomrule
\end{tabular}
\end{table*}

\subsection{PixelNet Results in The Regular Setting}
In this work, we are interested in studying PixelNet's performance by replacing the itemID embeddings in IDNet with a trainable vision encoder network.
We show the results of various PixelNet on Pixel200K  in Table~\ref{tab:compare}.  Some  results on Pixel1M and Pixel8M can be seen in Appendix Table~\ref{tab:benchmark_large}. 
For most experiments, we choose  two types of vision encoders for PixelNet that are widely used but not excessively large:  CLIP-ResNet50\footnote{see https://github.com/openai/CLIP}(denote as RN50) and CLIP-ViT-B\footnote{https://huggingface.co/openai/clip-vit-base-patch32}(denote as ViT).
We fine-tune their last two layers, as we notice that it is comparable to fine-tuning all parameters.

As shown, the comparison between PixelNet and IDNet exhibits significant differences in performance with non-sequential backbones (such as MF, FM, and DSSM) and sequential backbones. When utilizing  non-sequential architectures and corresponding training approaches,  PixelNet produces notably inferior results compared to their IDNet counterparts (Table~\ref{tab:benchmark} vs. ~\ref{tab:compare}). By contrast,  PixelNet shows  competitive or superior results (also see Figure~\ref{imeben}) over  IDNet when  using a sequential backbone.
The latter result is   surprising since  most of the prior literature shows that representing items  using purely visual features can outperform IDNet only in the very cold start setting~\cite{yuan2023go,du2020learn}. 
The superior results of PixelNet in the regular (with both warm and cold item) setting imply a significant advancement in the field  and potentially suggests that in the near future, PixelNet has the opportunity to considerably surpass IDNet by using more advanced image encoders from the computer vision field. If this happens, IDNet's dominance for visual content-driven recommendation would be substantially challenged, as PixelNet offers obvious advantages in other  scenarios, such as cold start and transfer learning recommendation settings (see below).

Furthermore, it is still unclear why PixelNet is effective with a sequential recommendation backbone but produces poor results, substantially worse than IDNet, with the classical MF backbone. It is plausible that learning a large image encoder is more challenging than learning a simple ID embedding vector, and therefore requires a more robust training approach --- for example, the autoregressive training mode of $\mathbf{i_1}, \mathbf{i_2} ... \mathbf{i_{n-1}} \rightarrow \mathbf{i_2}, \mathbf{i_3}...\mathbf{i_n}$ or $\mathbf{i_1}, \mathbf{i_2} ... \mathbf{i_{n-1}} \rightarrow \mathbf{i_n}$ (plus data augmentation) is generally  more effective than  the $\mathbf{u} \rightarrow\mathbf{i}$ mode, as shown in Table~\ref{tab:dataprocess}. These findings suggest that although PixelNet should theoretically work by replacing IDNet's itemID embeddings with image encoders, its actual performance may be significantly influenced by the specific recommendation backbone network and training approach used. 
 
	\begin{figure}
		\centering
		\subfigure[Non-sequential Recommender]{
			\begin{minipage}[t]{0.24\textwidth}
				\includegraphics[width=1\textwidth]{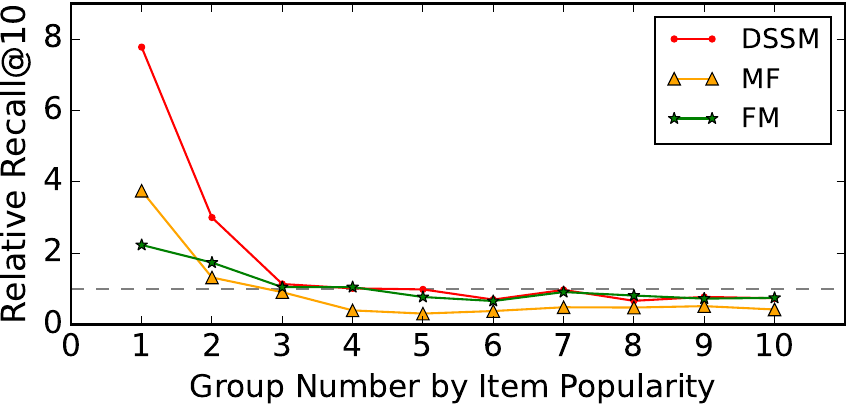}
				
			\end{minipage}
			
		}
  \hspace{-0.4cm} 
		\subfigure[Sequential Recommender]{
			\begin{minipage}[t]{0.23\textwidth}
				\includegraphics[width=1\textwidth]{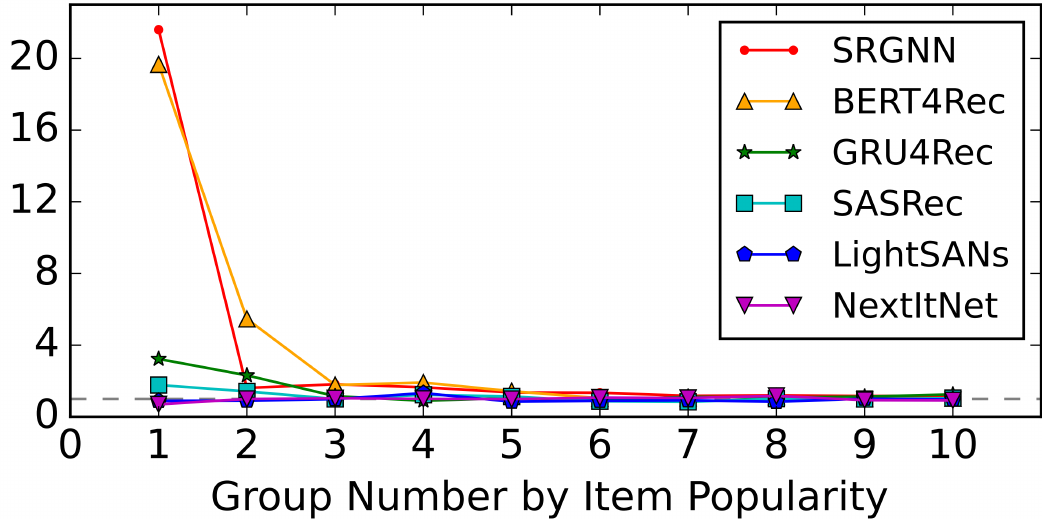}			
			\end{minipage}
                \label{fig:cold_se}
		}
	\caption{The y-axis represents the relative improvement on recall@10, which is the accuracy of PixelNet (with ViT as item encoder) divided by that of IDNet. Larger group number on the x-axis means that items in this group are more popular.}
	\label{fig:cold}
\end{figure}
\begin{figure}[t]
	\centering
	\includegraphics[width=1\linewidth]{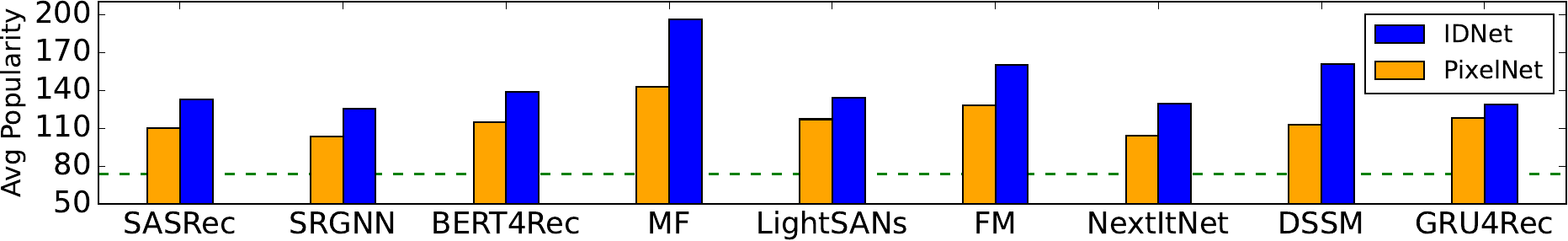}
	\caption{The y-axis is the average popularity of top-1 recommended items (i.e. the one with the highest score) for all users in the testing set. 
 The green dashed line is the average popularity of the target items in the testing set.}
	\label{fig:pop}	
\end{figure}
\subsection{Cold \& Warm Item Recommendation Setting }
It is recognized that utilizing (visual) content features can be a helpful approach in addressing the challenges of the cold start item problem.
To verify this on PixelRec, we denote item popularity by its number of occurrences in the training set, and sort the test data (200,000 user-item interactions, one item per user) from cold to hot according to its target item popularity. We then divide the test data into equally-sized groups, each containing 20,000 users. We compute the  improvement ratio on Recall@10 for each PixelNet over its IDNet counterpart. The results are presented in Figure~\ref{fig:cold}. It is evident that PixelNet, which includes DSSM, MF, FM, SRGNN, BERT4Rec, GRU4Rec, and SASRec, outperforms IDNet in relatively cold scenarios, especially the top two groups. Among these, MF, DSSM, BERT4Rec, and SRGNN achieve several-fold improvements for cold items. 
The findings demonstrate that PixelNet can effectively alleviate the cold-item problem on our PixelRec dataset. Regarding warm item recommendation, PixelNet with MF and DSSM backbones  performs significantly worse than IDNet.
In contrast, PixelNet with a sequential network backbone can perform comparably to IDNet in non-cold and even warm item settings (e.g., groups 8, 9, and 10 in Figure~\ref{fig:cold_se}.

We also compute the average popularity of the top-1 ranked items by both PixelNet and IDNet, as shown in Figure~\ref{fig:pop}. It indicates that the top-1 ranked items of PixelNet are  generally less popular than IDNet. This is a desirable property for a RS algorithm when competitive results are obtained, but more items are recommended from the non-head items.

\subsection{Cross-Platform Recommendation} 
As previously stated, PixelNet has a natural advantage in transfer learning tasks since it enables transfer between any two scenarios by acquiring general representations from image pixels.
We note that several recent works have begun  investigating  recommender system on text data~\cite{ding2021zero,hou2022towards,hou2022learning,shin2021one4all}, but few   \textit{peer-reviewed}  literature  have examined  transfer learning exclusively using raw visual features.

We report some preliminary results in Table ~\ref{tab:transfer}, since more convincing conclusions should be drawn based on multiple downstream datasets (beyond the focus of this paper). 
As we can see both SASRec and BERT4Rec have been improved by pre-training on the source dataset beforehand. For example,
recall@10 of SASRec using ResNet50 as image encoder increases from 10.959\% to 13.261\% by pre-training on PixelRec. Our results show that PixelRec can be a very good dataset as a pre-training dataset. We hope that PixelRec and our initial results will inspire new research in this direction.

\subsection{Benchmarking Vision Encoders}

	\begin{table}[t]
		\centering	
		\caption{Cross platform recommendation results(\%) on the target dataset (Tencent-News). `withPT' represents training the model initialized with pre-trained weights from the source dataset (Pixel200K), `noPT' means training the model without pre-training on the source recommendation dataset. We set the maximum training epoch to 120, R@10, N@10 is short for Recall@10, NDCG@10.}
		\setlength{\tabcolsep}{7pt}
		\label{tab:transfer}

		\begin{tabularx}{\linewidth}{l c l l l l }
			\toprule
			\multirow{2}{*}{Model} & \multirow{2}{*}{ImgEnc}&\multicolumn{2}{c}{noPT}& \multicolumn{2}{c}{withPT}\\
			\cmidrule(r){3-4} \cmidrule(r){5-6} 
			&& R@10 & N@10& R@10 & N@10 \\
			
			\midrule
			SASRec & RN50 & 10.959 &6.213 &13.261 & 7.417\\
			SASRec & ViT & 12.172&6.554& 13.009 & 7.239 \\

			BERT4Rec & RN50& 11.644 &6.069 &13.123 & 7.034\\
			BERT4Rec & ViT& 11.590 &6.091 & 12.931 & 7.245\\		
			\bottomrule
				\end{tabularx}
	\end{table}	
	
\begin{figure}
		\centering
		\subfigure[]{
			\begin{minipage}[t]{0.48\textwidth}
				\centering
				\includegraphics[width=.95\textwidth]{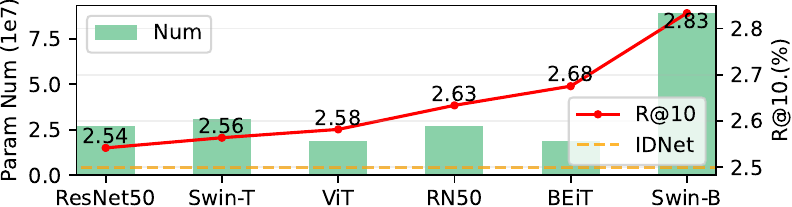}
				\label{fig:vision}
			\end{minipage}		
		}
		\subfigure[]{
			\begin{minipage}[t]{0.48\textwidth}
				\centering
				\includegraphics[width=.95\textwidth]{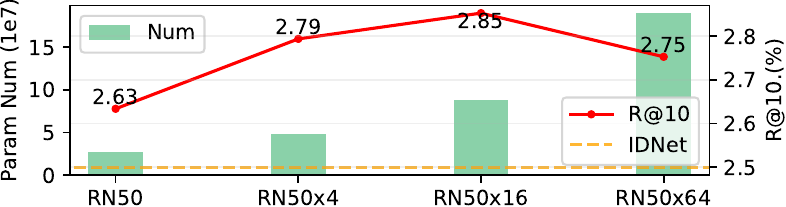}
				\label{fig:vision2}	
			\end{minipage}
         \label{fig:visionscale}
		}
		\caption{Benchmark Image Encoders. The dashed yellow line is the accuracy of IDNet. The green bar chart is the number of trainable parameters. The red line chart is the recall@10. }
		\label{imeben}
\end{figure}

\begin{figure}[t]
	\centering
	\includegraphics[width=.95\linewidth]{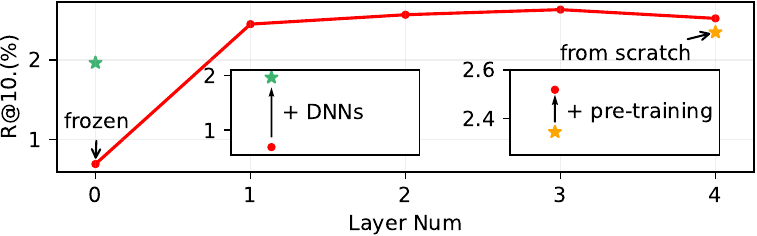}
	\caption{Joint training vs. the frozen-feature paradigm.  We use Swin-T for evaluation, consisting of  four author-defined layers in total.}
	\label{fig:layer}
	
\end{figure}

The effects of recommendation backbones (or user encoders) have been evaluated in Table ~\ref{tab:compare}, where we utilize  ViT and RN50 as image encoders.
Here we evaluate more image encoders  borrowed from the CV community. Given
the SASRec  backbone often provides state-of-the-art performance, we conduct exploratory experiments based on it.

We first evaluate six different image encoders, including ViT, RN50, ResNet50\footnote{\url{https://download.pytorch.org/models/resnet50-19c8e357.pth}}, BEiT\footnote{\url{https://huggingface.co/microsoft/beit-base-patch16-224}}~\cite{DBLP:conf/iclr/Bao0PW22} and Swin-T\footnote{\url{https://huggingface.co/microsoft/swin-tiny-patch4-window7-224}}, Swin-B\footnote{\url{https://huggingface.co/microsoft/swin-base-patch4-window7-224}} (the tiny/base version of Swin Transformer~\cite{liu2021swin}). Among them, ViT, Swin Transformer and BEiT are based on the MHSA or Transformer backbone, while others are based on the CNN network. 
Again, without special mention, all image encoders are fine-tuned for the last two layers. Then we evaluate the impact   by increasing the model size of RN50, including RN50, RN50x4, RN50x16, and RN50x64 (which correspond to roughly 4x, 16x, 64x the compute of a ResNet50, see ~\cite{radford2021learning}). Note that due to the large model size and GPU memory issue, only the last layer is optimized, 
but for RN50 we keep previous results. 
The results are shown in Figure~\ref{imeben}.

The results demonstrate that PixelNet combined with Swin-B and RN50x16 achieves the highest performance, with a recall@10 rate greater than 2.8\%. In contrast, when using a typical ResNet50 model, PixelNet's performance is relatively poor, even falling below 2.6\%.
On the other hand, PixelNet with Swin-B and RN50x16 comes with a higher computational cost due to the larger size of these models. Upon examining Figure~\ref{fig:visionscale}, we observe that larger image encoders do lead to improved performance, but only up to a certain point. Based on previous research, we conclude that both recommendation architectures  and image encoders play important roles in the effectiveness of PixelNet.

\subsection{Frozen Features vs. Joint Training}
\label{E2EEvaluation}
 
One of the key features of PixelRec is its capacity to facilitate the learning of recommendation models from raw image pixels. In other words, we can perform  joint learning of both the recommendation architecture and image encoders. 
Here, we assess whether joint learning  is more effective than the pre-extracted frozen visual features on PixelRec.
Figure~\ref{fig:layer} shows the comparison of frozen feature and joint training approach by fine-tuning different layers of image encoders. First, we freeze all layers and then gradually increase the number of fine-tuned layers from top to bottom. It can be seen that PixelNet has been greatly improved through parameter adaptation, even  fine-tuning only one layer/block. 
The optimal results are obtained by fine-tuning two or three layers.  In addition, we also show the results by gradually adding DNN  layers  on the frozen features and report the best results with seven layers (the green star in Figure~\ref{fig:layer}). Clearly, the accuracy improves, but is still noticeably  lower than fine-tuning image encoders. 
This finding suggests that the visual features learned in computer vision tasks are not sufficiently general, as a linear regression layer would yield the optimal outcomes if the features were comprehensive enough to encompass all elements of user preference. Finally, we also report results on PixelNet using a randomly initialized image encoder (the yellow star in Figure~\ref{fig:layer}). Undoubtedly, pre-trained parameters in the image encoder learned from  computer vision tasks  aid in achieving  better results from 2.34\% to 2.53\%.

\section{Conclusion}
In this paper, we introduced PixelRec, a very large content-driven image recommendation dataset that includes raw pixels. We benchmarked multiple recommendation paradigms, backbones, and item encoders on PixelRec. Our comparative analysis revealed the effectiveness of the PixelNet paradigm, which solely relies on raw image pixels to represent items, highlighting the crucial role of raw image pixel features in PixelRec. On the other hand, while PixelNet has many unique advantages, its typical training method is still costly. Nevertheless, we are optimistic that with PixelRec, researchers will be able to develop more advanced and practical visual recommendation algorithms, improve PixelNet, and even create new paradigms beyond PixelNet.

\vspace{12em}

\appendix

\section{Appendix}

\begin{table}[ht]
	\centering

		\centering
		\caption{Addtional results(\%) on Pixel1M and Pixel8M. MF, FM, SASRec, and BERT4Rec are IDNet.
			$\text{SASRec}_\text{vit}$ means SASRec uses the ViT as an image encoder.}
		\label{tab:benchmark_large}
		
		\setlength{\tabcolsep}{7pt}
		\begin{tabular}{ll c c c c }
			\toprule
			&Method  & R@5 &  N@5 & R@10 &  N@10    \\
			\midrule
			\multirow{8}{*}{Pixel1M}&MF & 0.619 & 0.381 & 1.102 & 0.536\\
			&FM & 1.058 & 0.662 & 1.759 & 0.887\\ 
			&SASRec & 2.583 & 1.688 & 4.004 & 2.144 \\
			&BERT4Rec & 2.563 & 1.655 & 4.062 & 2.137 \\
			\cmidrule(r){2-6}
			&VBPR& 0.437 & 0.268 & 0.773 & 0.375 \\
			\cmidrule(r){2-6}
			&$\text{SASRec}_\text{vit}$ & 2.765 & 1.777 & 4.402 & 2.303 \\
			&$\text{BERT4Rec}_\text{vit}$ & 2.727 & 1.753 & 4.343 & 2.272  \\	
			\midrule
			\multirow{2}{*}{Pixel8M} 

   &$\text{SASRec}_\text{vit}$ & 2.354 & 1.544 & 3.589 & 1.941 \\
			&$\text{BERT4Rec}_\text{vit}$ & 2.051 & 1.332 & 3.178 & 1.694 \\
			\bottomrule
		\end{tabular}
		\end{table}

	\begin{table}[h]
		\centering
		
		\caption{Training cost of MF and SASRec. \#Param: number of tunable parameters. Time/E: averaged training time per epoch  on Pixel200K  (most models below require about 100 epochs to converge).
			We measure Time/E on NVIDIA V100 (32G memory). Item encoder here is either ID-embedding (ID-Emb, or image encoder with full parameter fine-tuning (e.g. RN50-FT) or fine-tuning the last 2 layers (e.g. RN50-2T). 
			BS is the batch size for computing Time/E. 64(8) means 8 V100 GPUs are used and on each GPU the batch size is set to 64. 
   Taking SASRec + ViT-FT as an example,  the overall training time for Pixel200K on 8 V100 GPUs is about 85 hours.
		}

		\label{tab:time_compare}
		\setlength{\tabcolsep}{8pt}
  \begin{tabular}{lcrcr}
			\toprule
			Model & Item Encoder  & \#Param& Time/E & BS   \\
			\midrule
			\multirow{5}{*}{MF} & ID-Emb & 1213M &  	 0.25h & 512(1)\\
			& RN50-2T & 845M &  2.13h &  64(8)\\
			& ViT-2T & 836M  &1.70h & 64(8)\\
			& RN50-FT & 861M &  2.67h &  64(8)\\
			& ViT-FT & 909M  &2.26h & 64(8)\\
			
			\midrule
			\multirow{5}{*}{SASRec} & ID-Emb &53M & 0.03h & 64(1)\\
			& RN50-2T &26M & 0.75h & 8(8) \\
			& ViT-2T & 18M&  0.40h & 8(8)\\ 
			& RN50-FT &43M & 1.22h & 8(8)\\
			& ViT-FT & 92M&  0.85h & 8(8) \\ 
			\bottomrule
		\end{tabular}
	
\end{table}

\newpage
\normalem
\bibliographystyle{ACM-Reference-Format}
\bibliography{sample-base}

\end{document}